%% file: ms.tex
\documentstyle[emulateapj]{article}

\lefthead{Drake}
\righthead{Planetary Transits}

\begin{document}
\title{On the Selection of Photometric Planetary Transits}

\author{A.J. Drake\altaffilmark{1,2,3}}

\altaffiltext{1}{Dept. of Astrophysical Sciences, Princeton University, Princeton, NJ 08544}
\altaffiltext{2}{Depto. de Astronomia, P. Universidad Catolica, Casilla 104, Santiago 22, Chile}
\altaffiltext{3}{Lawrence Livermore National Laboratory, Livermore, CA 94550}

\begin{abstract}
  
  We present a new method for differentiating between planetary transits and
  eclipsing binaries based on the presence of the ellipsoidal light
  variations.  These variations can be used to detect stellar secondaries
  with masses $\rm \sim 0.2 M_{\sun}$ orbiting sun-like stars at a
  photometric accuracy level which has already been achieved in transit
  surveys. By removing candidates exhibiting this effect it is possible to
  greatly reduce the number of objects requiring spectroscopic follow up
  with large telescopes.  Unlike the usual candidate selection method, which
  are primarily based on the estimated radius of the orbiting object, this
  technique is not biased against bona-fide planets and brown dwarfs with
  large radii, because the amplitude of the effect depends on the transiting
  object's mass and orbital distance.  In many binary systems, where a candidate
  planetary transit is actually due to the partial eclipse of two normal stars,
  the presence of flux variations due to the gravity darkening effect will
  show the true nature of these systems.
  
  We show that many of the recent OGLE-III photometric transit candidates
  exhibit the presence of significant variations in their light curves and
  are likely to be due to stellar secondaries. We find that the light curves
  of white dwarf transits will generally not mimic those of small planets
  because of significant gravitationally induced flux variations.
  
  We discuss the relative merits of methods used to detect transit
  candidates which are due to stellar blends rather than planets.  We
  outline how photometric observations taken in two bands can be used to
  detect the presence of stellar blends.

\end{abstract}

\section{Introduction}

In recent years the discoveries of large numbers of planets via high
precession radial velocity studies (Marcy \& Bulter 2000) has fostered a
surge of activity aimed at the discovery of low mass companions to nearby
stars. However, with the radial velocity technique alone it is not possible
to uniquely determine the mass of a planetary candidate since the orbital
inclination ($i$) is poorly determined.  The uncertainty in $i$ leads to a
$sin(i)$ degeneracy in the mass of the orbiting object. In contrast,
planetary transit searches measure light curves which put strong constraints
on the inclinations of the planetary orbits.  This information can be combined with radial
velocity measurements to determine a transiting object's mass and mean
density.  Furthermore, once a transiting planet or brown dwarf has been
confirmed by radial velocity measurements, additional high accuracy spectral
and photometric observations make it possible to detect the presence of
specific chemical elements in a transiting planet's atmosphere (Seager \&
Sasselov 2000, Brown 2001, Charbonneau et al.~2002).

The radial velocity survey based discovery of a planetary companion to
HD~209458 and the subsequent discovery of a planetary transit (Henry et al.~
2000, Charbonneau et al.~2000, Mazeh et al.~2000) has led to a great deal of
interest in the detection of extra-solar planets via photometric transits.
The planetary object, HD~209458b, has a mass of 0.69M$_{\rm J}$, a radius of
1.35R$_{\rm J}$, and an orbital distance of 0.047AU (Cody \& Sasselov 2002).
A number of other extra-solar planets have been discovered with small
separations from their parent stars in radial velocity surveys (51-Peg-b,
$\tau$-Boo-b, HD~187123b, etc.), but no photometric transits have been
observed for these.  

Within the last year planetary transit candidates have been discovered in
the data from the OLGE-III (Udalski et al.~2002a,b) and the EXPLORE
(Mall\'en-Ornelas et al. 2002) projects as well as the Vulcan 
campaign (Jenkins et al.~2002).
However, when considering whether a photometric transit is due to a planet
or a small star, such as a late M-dwarf, it is necessary to know the mass of
the transiting object. This mass can be determined with radial velocity
measurements by using large telescopes with high resolution spectrographs
(such as VLT or Keck).  However, the determination of the mass of a single
object may require a number of observations taken on multiple nights.  As
there are currently more than 60 planetary transit candidates and many
hundreds more expected from space missions and other searches, it will soon
be impractical to measure the radial velocity profiles of all candidates.

A method of selecting planetary transit candidates has been put forward by
Seager \& Mall\'en-Ornelas (2002). Among other things, this selection favors
planets that have circular orbits, unblended parent stars and produce flat
bottomed eclipse shapes.  However, it is clear that any of these criteria may be
restrictive against bona-fide planet transits.  Circular orbits seem
probable for transiting planets since they are strongly influenced by
gravitational tidal forces due to their close proximity to their parent
stars.  However, this may not always be true since significant
eccentricity has been observed among the planets found in radial velocity
surveys (Udry et al.~2002).  The second selection criteria can also fail to
produce all good candidates because in some cases (such as OGLE-III) dense
stellar fields are monitored to maximize the planet detection efficiency.
Furthermore, many blends are likely to occur because of multiple star
systems (Mall\'en-Ornelas et al.  2002).  The selection of flat bottomed
light curves excludes planets transiting near stellar limbs.  These may not
exhibit any flat regions whatsoever.  Although the number of transits
expected at these inclinations is generally small, the exact fraction of limb
transits is strongly dependent on the as-yet-unknown sizes of tightly
orbiting planets.  Furthermore, the effect of limb darkening is strong in
photometric bands blueward of I-band. Thus, a large fraction of bona-fide
planets may not exhibit flat bottomed eclipses.  The effect of limb
darkening on the shape of planetary transits is clear from the HST data of
planetary transit HD~209458 (Brown et al.~2001).

The selection of good planetary candidates can be improved if one knows the
size of the parent star $R_{\star}$.  This can be estimated with some accuracy by
determining the spectral type of the parent star. Since the fractional drop
in flux during the eclipse, $\Delta F$, is related to the stellar and
planetary radii by,

\begin{equation}
\Delta F = \left(\frac{R_{p}}{R_{\star}}\right)^{2}\!\!,
\end{equation}

\noindent
a determination of the candidate planet's radius $R_{p}$ can be obtained.
However, limb darkening changes the measured $\Delta F$ by a small amount
which depends on the stellar type and the passband used.

The evolutionary theory of isolated planets by Baraffe et al.~(1998, 2002)
predicts that evolved extra-solar planets should have radii similar to that
of Jupiter.  Dreizler et al.~(2002) used these models to select the best
transit candidates from Udalski et al.~(2002a,b), for follow-up with radial
velocity measurements based on their sizes.  However, currently there are
little empirical data about the actual radii of gaseous planets at small
separations from stars.  For example, presently the only extra-solar planet
with a measured radius is HD~209458b and as Dreizler et al.~(2002) conceded,
their adopted model is inconsistent with HD~209458b.  It would seem that
evolutionary models of isolated planets are poor representations of real
planets.  The giant extra-solar planet models of Burrows et al.~(2000),
Bodenheimer et al.~(2001) and Guillot \& Showman (2002) suggest that large
extra-solar planets can exist at small distances from main-sequence stars.
Guillot \& Showman (2002) point out that, if only 1\% of the flux incident
on HD~209458b from its parent star was transformed to kinetic energy in the
planetary atmosphere, a planet can maintain its size.  In such
models the incident radiation suppresses the typical energy loss by
radiative diffusion and hence the contraction of such planets.  As many
planets have been found at small distances from their host stars in radial
velocity surveys, it seems that gaseous planets at small orbits could have
larger radii than HD~209458b.  Furthermore, it seems possible that
transiting brown dwarfs may also have large radii.  In light of the current
level of uncertainty about the radii of extra-solar planets and brown dwarfs
in tight orbits, it would be useful to select candidates based on tracers
for the transiting objects' masses as well as their sizes. One tracer for the
mass of the transiting companion in tight binary systems is the presence
of gravity darkening effects in eclipsing binary light curves.

In the next section we outline the nature of the ellipsoidal variations
observed in binary light curves due to gravity darkening.  In \S 3, we show
how the effect can be used to select bona-fide planet/brown dwarf
candidates. In \S 4, we test some public photometric transit data for the
presence of the gravity darkening effect. Following this in \S 5, we
discuss the problems with planetary transit selection due to photometric
blending. A partial solution to the blending problem will be discussed in \S
6, followed by concluding remarks.

\section{Gravity darkening}

When two stars exist in a tight orbit, their gravitational potential
stretches the envelopes of the stars into ellipsoids (von Zeipel 1924).
These distortions leads to sinusoidal modulations of the kind 
observed in the light curves of stars such as W UMa 
binary systems (Kitamura \& Nakamura 1988).  When these binary systems
rotate their brightnesses vary depending on the observed luminosity and
cross section of the stars.  The temperature of the surface elements is seen
to act in proportion to the effective gravity, such that

\begin{equation}
T_{eff}^{4} \propto g^{\beta_{1}}.
\end{equation}

It is clear that the extent of the gravity darkening effect observed depends
on the exponent $\beta_1$. This parameter varies strongly between radiative stars 
(where it is 1) and convective stars (where it is typically around 0.25).  The
exact value of $\beta_1$ varies with the stellar mass and has been modeled
for a range of stellar parameters by Alencar \& Vaz (1997) and Claret (1998, 2000).
However, there is only a small amount of empirical data available, 
such as Rafert \& Twigg (1980).

The sinusoidal modulation due to ellipsoidal geometry of the stars in a
binary system varies at twice the orbital frequency.  Since the distortion
of the primary is independent of whether or not the secondary is luminous,
it is an ideal tracer for an unseen massive companion to a luminous star
(Beech 1985).  Furthermore, when both stars have similar luminosities the
reflection effect can also be observed in the light curves.  The ellipsoidal
effect can be observed in binary systems even if an eclipse is not observed,
because the ellipsoidal effect can be observed over a larger range of
orbital inclinations than a transit.  In the next section, we will show
that, although accurate values of $\beta_1$ are yet to be determined
empirically for many stellar types, even the slightest signal of a
modulation can point to a secondary object with a significant mass.

\section{Selecting planetary candidates}

Typical planetary transits are expected to last a few hours for orbital
periods of a few days. These dips in flux during the transit will be $ \la
1\%$. Therefore, the detection of a planet transit requires a well sampled
light curves with high accuracy data.  Transit detections can be made by
either making frequent observations of candidate stars (dedicated transit
surveys), or by folding light curves and searching for significant periods
in data taken less frequently over a longer time (such as microlensing
surveys).  Because planet transits are short, roughly 95\% of the photometry
points taken in a survey will occur outside the transit region for any given
light curve.  These measurements are of little use to the determination of a
planet's size or shape. However, these data make it possible to measure
variations in the baseline of the light curve to much greater accuracy than
the transit depth.

The OGLE-III transit search data set consists of 800 photometry points
(Udalski et al.~2002a,b) with an accuracy of better than 1.5\%. A simple
Poisson argument would suggest that at a 3 sigma level, a modulation
amplitude of $\sim 1.6 \times 10^{-3}$ magnitudes should be measurable. For
the Explore-I project the best candidates ($15 \la I \la 17$) have 1600
photometry points with better than 0.5\% photometry. In this case, we expect
that a variation should be detectable at the level of $\sim 3.75 \times
10^{-4}$ magnitudes.  However, in reality it is likely that there are
systematic contributions to the uncertainty in the photometry points.
Systematic errors can occur due to observations of a field being taken
nightly at a similar airmass range.  The observed scatter will not behave in
a Poissonian way, so the actual measured baseline uncertainties will always
exceed this limit.  Indeed, many of the transit light curves presented by
the OGLE-III project appear to exhibit time dependent fluctuations which are
either due to systematic noise contributions or real variability of the
parent stars.

\begin{figure*}
\plotone{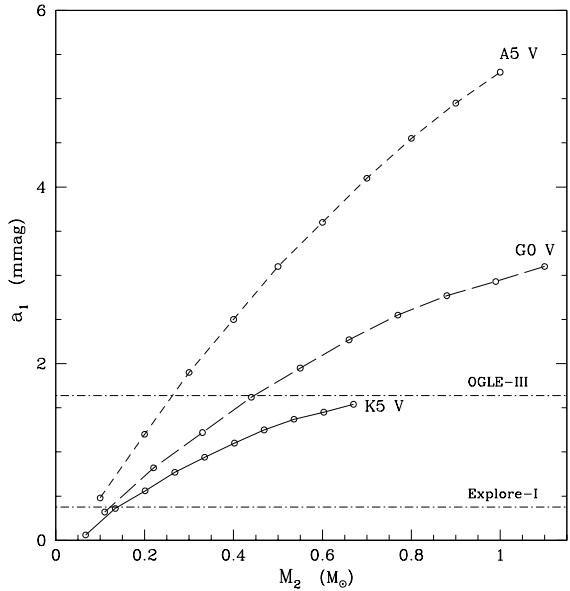}  
\figcaption{
  The amplitude a$_{1}$ of light curve baseline luminosity variations due
  to the gravity darkening effect, as a function of secondary masses
  M$_{2}$.  The displayed curves are for three types of primary stars: an A5
  V spectral type (assuming 2.0M$_{\sun}$, 8200K, 1.7R$_{\sun}$), a G0 V
  type (assuming 1.1M$_{\sun}$, 6000K, 1.1R$_{\sun}$), and a K5 V type
  (assuming 0.67M$_{\sun}$, 4200K, 0.72R$_{\sun}$), each with a 2 day period
  orbiting secondary. The two dash-dotted lines show the 3-sigma detection
  significance levels for ellipsoidal modulations detected in the OGLE-III
  data and Explore-I data as noted in the text.\label{dark}}
\end{figure*}

\placefigure{dark}

In Figure \ref{dark} we present the modulation amplitudes expected due to
the gravity darkening effect.  These values have been determined using the
{\em Nightfall}\footnote{See
  http://www.lsw.uni-heidelberg.de/$\sim$rwichman/Nightfall.html for
  details} eclipsing binary analysis program. This program can be used to
model and fit the shapes of eclipsing binary light curves, line profiles and
radial velocities, and includes effects such as limb darkening, star spots,
reflections and third light .  The values for the gravity darkening exponent
$\beta_1$ are taken from the non-grey models of Alencar \& Vaz (1997).  For
the curves in this particular example we have taken an orbital period of 2
days. This is consistent with a large number of the OGLE-III candidates.  We
expect that the spectral types of primary stars will generally be within the
range of types shown on Figure 1.  For example, the OGLE-III candidates
measured by Dreizler et al.~(2002) have spectral types ranging from A3 V to
K4 V. Stars with spectral types later than K5 V are very faint and are,
therefore, difficult to photometer accurately, while types earlier than A5 V
(and giant types) will only have small planetary transit signals due to their
sizes.

Current planetary transit surveys are attempting to find planets and brown
dwarfs with masses in the range $\rm M_{2} < 0.08 M_{\sun}$. From Figure 1
we see that (for most types of primary stars), the detection of modulations
greater than $> 0.5$ mmag in amplitude is sufficient to reject candidates
for orbital periods longer than 2 days.  For larger orbital periods, the
separations of the stars and candidate planets are greater and the gravity
induced modulations are smaller.  Thus, the presence of measurable
modulations at longer periods provides even stronger evidence that transiting
objects have significant masses.  For the most accurately measured stars it
may be possible to limit the secondary's mass to within the brown dwarf mass
range $\rm < 0.08M_{\sun}$ using photometry alone. This will be very useful
for reducing the number of possible planetary candidates in space missions
such as Kepler.

The transits of giant-type stars by main-sequence stars can be distinguished
from those of planets eclipsing main-sequence stars, by using the parent
star's stellar density derived from the transit parameters (Seager \&
Mall\'en-Ornelas 2002). These stars can also be distinguished 
with spectra or multi-band photometry (Bessell \& Brett 1988).
Furthermore, the ellipsoidal effect is larger for giant stars than
main-sequence stars since they are more readily distorted due
to their low surface gravities.

If a candidate planetary transit is due to the partial eclipse of two normal
stars, the observed gravity darkening effect may be large even though the
observed eclipse may be very small.  In some cases an observed ellipsoidal
effect could also be due to the presence of additional massive object which
is not the transiting object. Such a third object would have an orbital period
different from that of the transiting object and thus would appear as a
source of correlated noise in the phased transit light curve. In other
cases a transit light curve may exhibit sinusoidal modulations due to
effects other than the ellipsoidal effect. For instance, one can imagine
cases where an observed modulation was due to the concentrations of star spots on
one hemisphere of a star.  However, this is unlikely to properly mimic the
gravity darkening effect since this would require that the rotational period
of the star was exactly twice the orbital period of a planet.  
Furthermore, the number and distribution of spots is likely to change over time.

Because there is some uncertainty in the value of $\beta_1$, the exact
amplitudes presented in Figure \ref{dark} are also quite uncertain.  In any
case, it seems prudent to reject candidate planetary transits exhibiting
sinusoidal modulations whether they are due to a massive transiting
object or some other phenomena.

\section{Examination of the OGLE-III data}

To test whether the ellipsoidal effect can really be used to select planetary
transit candidates we retrieved the freely distributed OGLE-III transit data
from the OGLE-III archive\footnote{ftp://bulge.princeton.edu/ogle/ogle3/transits}.  
We phased these light curves with the most recently derived periods which are 
given on the OGLE-III web site\footnote{http://bulge.princeton.edu/$\sim$ogle}.
Of the OGLE-III transit candidates, numbers 43 to 46 do not have known
periods so could not be phased.  On examination of the transit candidate 
OGLE-TR-39 we found it to have, not only a strong sinusoidal modulation
but additional features visible at a period of 2.44565 days (three times 
the OGLE-III period). The presence of this additional signal would invalidate
any sinusoidal fit, so we did not analyze this candidate.
For each of the remaining set of light curves we subtracted the data during
the period of the transit.  Simply fitting the data with the inclusion of
the transit dip would clearly bias the results toward higher values.  In the 
absence of any ellipsoidal or other systematic effects, the data outside the
transit region should be constant.

Many of the OGLE-III transit light curves show the clear signs of systematic
trends with time. The exact origins of these features are unknown. However,
such features are common in the photometry of crowded fields with large
cameras. These occur because of blending between neighbouring stars, flat
fielding errors, inaccurate airmass corrections, changes in transmission
(due to atmospheric dust), etc.  To obtain a reliable estimate of whether
the effect of gravity is significant in any light curve, it is necessary to
have a good estimate of the photometric scatter.  In this analysis we
attempted to determine the significance of the ellipsoidal variations in the
presence of the real systematic noise.  To do this, firstly, we fitted the
phased light curves with the ellipsoidal modulation approximated by $\rm
a_{1}cos(2\phi)$ plus a constant (where $a_{1}$ is the gravity
darkening amplitude).  We note that Beech (1985) and others have shown that,
although additional sinusoidal terms exist, they are small in comparison to
the $\rm cos(2\phi)$ term.  The initial fits resulted in large $\chi^{2}$
values and very significant sinusoidal amplitudes because of the
underestimated uncertainties in the data points.  Next, we subtracted the
fitted values from the data to remove the gravity darkening trend.  Based on
a Gaussian uncertainty model, we determined the time averaged uncertainty in
the residuals for each light curve.  We scaled the error bars to match the
observed scatter and determined the parameter uncertainties in the fit
terms.  In many cases the fit $\chi^2_r$ value remained high because of real
variability structure within the light curves.  As an additional test of the
significance of the gravity darkening term we perform an F-distribution test
for the significance of this extra parameter. That is, we compared the
fit $\chi^{2}_{1}$ value for a constant baseline with the $\chi^{2}_{2}$
value for a constant term plus sinusoid.  The F-test statistic is given by

\begin{equation}
F = \frac{\chi^{2}_{1} - \chi^{2}_{2}}{\chi^{2}_{r2}},
\end{equation}

\noindent
where $\chi^{2}_{r2}$ is the reduced chi-square value of the fit with the
extra term. This statistic obeys the F-distribution for $\nu_1 = 1$ and
$\nu_2 = N-m$ ($N$ points, $m$ parameters) degrees of freedom. An
F-distribution value larger than 6.668 (for 750 data points) is expected in
only 1\% of experiments and larger than 10.91 for 0.1\%. The larger this
value the more significant the decease in the fit $\chi^{2}$ is with the
additional term.  In Table \ref{tab1} we present the fit values for the
OGLE-III transit candidates with amplitudes, $a_{1}$, greater than zero at a
3.5 sigma level and F-statistic values greater than 6.668.  We believe that
these transit candidates are the most likely to have stellar mass 
secondaries.

\begin{figure*}
\plotone{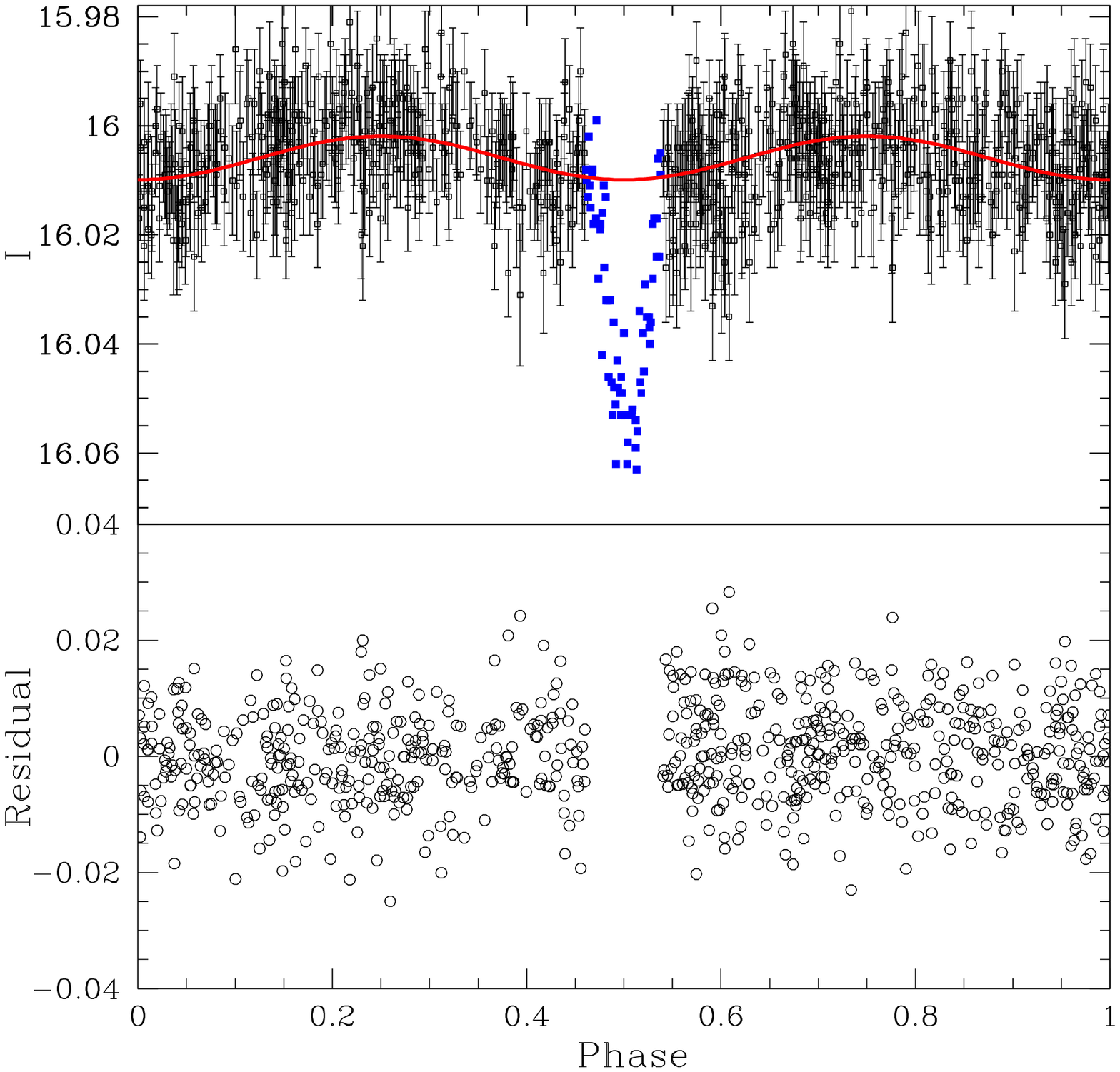}\hspace*{1.5cm}\plotone{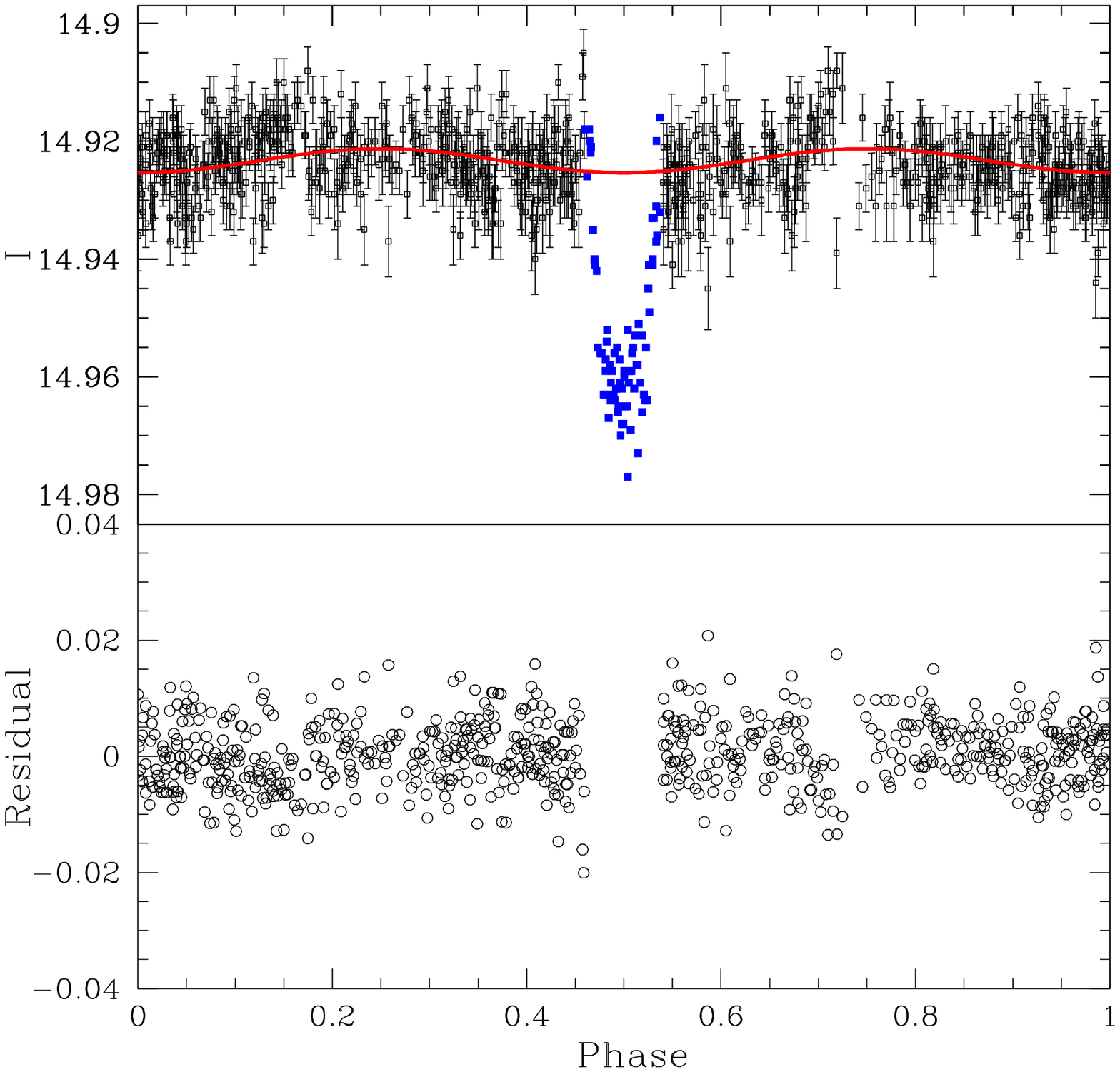}
\figcaption{The light curves and residuals for the transit candidates
  OGLE-TR-18 (left) and OGLE-TR-30 (right) are shown when fitted with a
  term of the form $\rm cos(2\phi)$. The filled square points
  were removed during the fit to prevent any bias due to the eclipse.\label{OGLE}
}
\end{figure*}

\placefigure{OGLE}

\placetable{tab1}

We note that the planetary transit candidate OGLE-TR-3 was selected by
Dreizler et al.~(2002) from the OGLE-III candidates as a likely planetary
transit. This object exhibits a sinusoidal modulation at the $4\sigma$
level.  The light curve also appears to exhibit the presence of
a second dip in the phased light curve, although this dip is not
very significant ($\sim 2\sigma$). The presence of a second dip in a binary
light curve is the clear sign that transiting object is luminous.  In Figure
\ref{OGLE} we present the fits to two of the OGLE-III planetary transit
candidates, OGLE-TR-18 and OGLE-TR-30. The presence of a sinusoidal
modulation is clearly seen in these light curves.

\section{Blending}

A common type of contaminant in searches for planetary transits are cases
where an ordinary eclipsing binary star is blended with one or more
additional stars.  The light curves of these candidates exhibit a dip in
flux where one of the stars is eclipsed. This may mimic a planetary transit
as only a small dip in the group flux is observed.  The additional flux
from the blended stars causes the size of eclipsed object to be
underestimated from the light curve.  As an example, take an M5 V spectral
type star eclipsing a K2 V type star.  Without any blended flux the complete
stellar eclipse would be too long and too deep to be due to a planet.
However, if these stars were blended with the light from an F5 V type star,
the transit time and the dip in flux would be consistent with a $\sim 1.3
R_{J}$ planet transiting the F5 type star near its limb.  In this example
we have assumed that the F5, K2 and M5 stars have the following typical
parameters, $\rm M_V =$ ($3.5$, $6.4$, $12.3$) and $\rm R_{\sun} =$ ($1.3$,
$1.1$, $0.798$), respectively.

Seager \& Mall\'en-Ornelas (2002) have shown that the percentage of blended
events can be derived statistically by using the ratio of the flat part of
the eclipse to the total transit time.  However, this statistical information
does not determine which of the light curves are blended in a given survey.
Alternatively, if a spectrum has been taken of the parent star, it is
possible to assume an approximate mass and radius based on the spectral type.
With the additional information it is possible to compare the density of
the parent star with that derived from transit parameters, timescales (total
and flat), depth, and orbital period (Seager \& Mall\'en-Ornelas 2002).
However, this process is complicated when the blended star is much brighter
than the stars involved in the eclipse.  In such cases, the measured spectra 
and density will be for the blended star rather than the eclipsing stars.

In principal it is also possible to determine the presence of blending
astrometrically since the location of the photo-center of the blended system
will vary from the location of the eclipsing stars.  The location of the
eclipsing system can be found by using difference image analysis (Alcock et
al.~1999).  This technique allows the determination of the photo-center of a
variable object or binary system by subtracting the constant flux component
(the blended baseline flux).  
An accurate location can be found by combining difference images taken at
times during the eclipse. The position of the eclipse is the location of 
the residual flux. 
A significant offset between the position of the residual and the
photo-center of the blended system can prove that candidate is blended
(Alcock et al.~2000, 2001).  However, in some cases the separation between the
blended centroid and the eclipse centroid will be indistinguishable. 
Furthermore, many surveys use very large pixel sizes to survey large areas of 
sky. In such cases astrometry is probably impractical.

The gravity darkening effect occurs whether or not a planetary transit
candidate is blended.  However, the amplitude of the observed ellipsoidal
light variations depends on the brightness of the eclipsing stars relative
to the additional flux component. In our example above, the ellipsoidal
modulation is very small for periods longer than a day and is washed out by
the flux from the much brighter blended star.  
If the blending is caused by an association of stars, the number of planets
is limited to the few stable orbital configurations in such systems.
Alternatively, if the additional blended flux comes from the transiting
object, such as when the planet candidate is due to a partial stellar
eclipse, ellipsoidal variations will be detectable with good photometry.

\section{Using colour information}

In section 5 we demonstrated how a blended stellar system can mimic a
planetary transit. We will now consider how colour information 
can be used to select bona-fide transits.
Our blending example consists of F5, K2 and M5 type stars.  An F5 V star has
a $(V-I)_{KC}$ colour of $\sim 0.53$ while a K2 V star has a colour of $\sim
0.98$.  The difference between the colour of these two stars will led to a
variation in the transit depth when observed in multiple passbands. 
For our test case the maximum observed transit depths will be $0.011$
magnitudes in $V$-band and $0.007$ magnitudes in $I$-band. Although this
difference is very small, it is a large fraction of the eclipse depth and 
is measurable at the accuracy levels achieved in current transit surveys.  

The colours of stars on the main-sequence vary in relation to their sizes.
The difference in the transit-depths observed in two passbands will
increase as the difference in the sizes of the blended stars increases.  
Therefore, as the amount of blending increases the fractional difference 
in observed transit depth will generally increase.
The variations in transit depth are most easily observed when observations are
taken at widely separated wavelengths.  With this in mind, it may well
be worth pursuing an observational search strategy where filters are swapped
between each observation.  Such a strategy can achieve a sampling
rate similar to a single colour survey (depending on the observation times and
passbands chosen), and gain additional information about colour evolution.

Photometric planetary transit searches require light curves exhibiting a single
eclipse since the planet does not contribute to the luminosity of the
system.  When the orbit of a binary is non-circular there are inclinations
where a partial eclipse can appear like a planetary transit.  In such cases
the orbital eccentricity can cause only a single eclipse to be observed.
The presence of a second eclipse is also significantly reduced when there are 
large differences in the luminosities of the stars in a binary.
Therefore, candidate planetary transits are also biased towards binary stars 
with large luminosity differences.
If both stars are on the mainsequence they are likely to have different 
colours. This colour information can be used to remove candidate planetary 
transits due to partial stellar eclipses.
The detection of a significant colour change during an eclipse is
evidence of blending.  Colour changes have been observed in searches for
planetary transits candidates in MACHO project data (Drake et al.~2003).

Testing planetary transit candidate light curves for colour changes during
eclipses may be the easiest and most robust way of finding which events are
due to blending.  In many cases, a few high signal-to-noise photometric
follow up observations, inside and outside the eclipse, will detect the
presence of blending. However, in some situations, radial velocity
measurements will still be necessary to determine whether or not blending is
responsible. It is clear that none of the techniques listed above can give a
clear indication of whether blending is present in all cases. Still, the
combination of these techniques should reduce the number of blends in
planetary transit searches.

\section{Conclusions and Discussion}

We have shown that the presence of modulations in the light curves of a
planetary transit candidates is likely to be due to the gravity darkening
effect.  This photometric information can be used as an effective way of
reducing the number of spurious candidates in current and future planet
searches.  Unlike the transit timescale, the gravity darkening modulations
are directly related to the secondary's mass.  Such a size-independent
relation is important because of current uncertainty about the sizes of
planets and brown dwarfs in close stellar orbits.

In many cases when very good photometry ($< 1\%$) is obtained, it is
possible to rule out transit candidates due to M-dwarfs to white dwarfs.  If
greater than 1000 highly accurate ($< 0.5\%$) photometry points are
obtained, it may be possible to select objects in the planet/brown dwarf
regime with photometry alone. The {\em Kepler} mission is expected to obtain
a photometric precision of 90 micromags (Sahu \& Gilliland 2002).  However,
radial velocity follow-up will still be necessary to determine accurate
masses of individual objects and thus to separate brown dwarfs from planets.

White dwarf secondaries are expected to mimic small extra-solar planets in
transit surveys due to the transit depths they will cause.
It has been shown by Marsh (2001) that microlensing of a transiting white
dwarf can cause a magnification during the transit.  It has also been
reported by Sahu \& Gilliland (2002) that the presence of this lensing can
be used to break the similarity between planetary and white dwarf transits.
However, as white dwarfs are expect to have masses around 0.6 $\rm
M_{\sun}$, the gravity darkening effect will be strong. Therefore, at
small orbital distances ($< 0.03 AU$) white dwarf transits will not appear
similar to planets.  At larger orbital distances microlensing becomes
important.  However, few transits are expected with long periods since the
transit probability decreases as $R_{*}/a$, where $R_{*}$ is the primary
star's radius and $a$ is the orbital separation.

In this work we have shown that many of the Udalski et al.~(2002a,b)
OGLE-III transit candidates exhibit the presence of ellipsoidal
modulations in their light curves with $> 3.5\sigma$ significance. In
the coming years many hundreds of planetary candidates are expected from
on-going transit surveys.  Such large numbers may necessitate the use of a
lower selection threshold. However, any selection is tentative without a clear
understanding of the systematic uncertainties in the light curves.

Udalski et al.~(2002a,b) selected main-sequence stars as targets under the
assumption that the transiting objects passed before the center part of
their parent stars.  Under this assumption it is possible to derive the
parent star's radius and mass from the transit light curve parameters using
the equations presented by Seager \& Mall\'en-Ornelas (2002).  However,
Udalski et al.~(2002a,b) instead assumed that all candidates had a mass of
$\rm 1 M _{\sun}$ and simply noted that there was a 
small scaling for parent star and planet with the true mass.  Although this
assertion is correct, the list of OGLE-III candidates contains many stars
with radii which are inconsistent with a standard main-sequence mass-radius
relationship. Since the radii of transiting objects are derived
from the sizes primary stars, these values are biased by the 
$\rm 1 M _{\sun}$ assumption.

When a bright star is blended with an eclipsing binary system the resultant
light curve can mimic the signal of a planet transiting the bright star.  In
such cases it may be difficult to separate bona-fide candidates from blends
with spectra or based on the presence of sinusoidal baseline variations.
However, it may be possible to detect the blend by measuring the depth of
the transit in more than one passband.

I wish to thank the members of OGLE-III project for making their data publicly
available.  I also wish to thank Gabriela Mallen-Ornelas and Sara Seager for
their comments which improved this paper.

\include{table1}

\end{document}

%% file: table1.tex
\begin{deluxetable}{lrccr}
\tablecaption{Transits Exhibiting Gravity Darkening.\label{tab1}}
\footnotesize
\tablewidth{0pt}
\tablehead{\colhead{ID} & \colhead{Amp} & \colhead{Err} & $\chi^{2}_{r}$ & \colhead{F}\\
& (mmag) & (mmag) &
}
\startdata
OGLE-TR-2   &   3.21 &  0.27 &  1.45 &   149.7 \\ 
OGLE-TR-3   &   1.49 &  0.37 &  0.88 &    18.1 \\ 
OGLE-TR-5   &   7.23 &  0.39 &  1.49 &   323.4 \\ 
OGLE-TR-6   &   1.86 &  0.46 &  1.21 &    16.4 \\ 
OGLE-TR-7   &   1.79 &  0.36 &  1.29 &    23.6 \\
OGLE-TR-13  &   1.46 &  0.25 &  1.51 &    35.8 \\ 
OGLE-TR-14  &   1.60 &  0.25 &  1.43 &    46.2 \\ 
OGLE-TR-16  &  14.58 &  0.29 &  1.69 &  2573.9 \\ 
OGLE-TR-18  &   3.99 &  0.44 &  1.01 &    85.2 \\ 
OGLE-TR-21  &   1.79 &  0.38 &  1.00 &    23.7 \\ 
OGLE-TR-24  &   2.23 &  0.64 &  3.00 &    16.4 \\ 
OGLE-TR-25  &   3.28 &  0.38 &  1.35 &    72.5 \\ 
OGLE-TR-27  &   7.28 &  0.57 &  1.28 &   171.2 \\ 
OGLE-TR-30  &   2.04 &  0.31 &  1.33 &    44.7 \\ 
OGLE-TR-31  &   3.68 &  0.29 &  1.59 &   172.9 \\ 
OGLE-TR-32  &   6.01 &  0.27 &  1.28 &   454.3 \\ 
OGLE-TR-40  &   0.97 &  0.26 &  1.42 &    12.2 \\ 
OGLE-TR-52  &   3.67 &  0.56 &  1.33 &    42.8 \\ 
OGLE-TR-57  &   5.92 &  0.44 &  1.01 &   194.0 \\ 
\enddata
\tablecomments{
Col. (1), OGLE-III object ID.
Cols. (2) \& (3) Fit amplitude and uncertainty. 
Col. (4) Reduced $\chi^{2}$ value of fit.
Col. (5) F-statistic value.
}
\end{deluxetable}

%% file: ms.bbl
\begin{thebibliography}{}

\bibitem[\protect\citeauthoryear{{Alcock} et~at.}{{Alcock} et~al.}{1999}]{Alcock99}
{Alcock}, C., et~al. 1999, ApJ, 521, 602
\bibitem[\protect\citeauthoryear{{Alcock} et~at.}{{Alcock} et~al.}{2000}]{Alcock00}
{Alcock}, C., et~al. 2000, ApJ, 541, 734
\bibitem[\protect\citeauthoryear{{Alcock} et~at.}{{Alcock} et~al.}{2001}]{Alcock01}
{Alcock}, C., et~al. 2001, ApJ, 552, 582
\bibitem[\protect\citeauthoryear{{Alencar}, \& {Vaz}}{{Alencar}, \& {Vaz}}{1997}]{Alencar97}
{Alencar} S.H.P. \& {Vaz}, L.P.R., 1997, A\&A, 326, 257 
\bibitem[\protect\citeauthoryear{{Baraffe} et~at.}{{Baraffe} et~at.}{1998}]{Baraffe98}
{Baraffe}, I., et~al., 1998, A\&A, 337, 403 
\bibitem[\protect\citeauthoryear{{Baraffe} et~at.}{{Baraffe} et~at.}{1998}]{Baraffe02}
{Baraffe}, I., et~al., 2002, A\&A, 382, 563
\bibitem[\protect\citeauthoryear{{Beech}}{{Beech}}{1985}]{Beech85}
{Beech}, M.,  1985, Ap\&SS, 117, 69
\bibitem[\protect\citeauthoryear{{Bessell}, \& {Brett}}{{Bessell}, \& {Brett}}{1988}]{Bessell88}
{Bessell}, M.S., \& {Brett}, J.M., PASP, 1988, 100, 1134
\bibitem[\protect\citeauthoryear{{Bodenheimer}, {Lin}, \& {Mardling}}{{Bodenheimer}, et~al.}{2001}]{Bod01}
{Bodenheimer}, P., {Lin}, D.N.C., \& {Mardling}, R.A., 2001, ApJ, 548, 466
\bibitem[\protect\citeauthoryear{{Brown}}{{Brown}}{2001}]{Brown01a}
{Brown}, T.M., 2001, ApJ, 553, 1006
\bibitem[\protect\citeauthoryear{{Brown} et~al.}{{Brown} et~al.}{2001}]{Brown01b}
{Brown}, T.M., et al., 2001, ApJ, 552, 699
\bibitem[\protect\citeauthoryear{{Burrows} et~al.}{{Burrows} et~al.}{2000}]{Burrows00}
{Burrows}, A., et~al., 2000, ApJ, 534, 97
\bibitem[\protect\citeauthoryear{{Charbonneau} et~al.}{{Charbonneau} et~al.}{2000}]{Charbonneau00}
{Charbonneau}, D., et~al., 2000, ApJ, 529, 45
\bibitem[\protect\citeauthoryear{{Charbonneau} et~al.}{{Charbonneau} et~al.}{2002}]{Charbonneau02}
{Charbonneau}, D., et~al., 2002, ApJ, 568, 377
\bibitem[\protect\citeauthoryear{{Claret}}{{Claret}}{1998}]{Claret98}
{Claret}, A., 1998, A\&AS, 131, 395
\bibitem[\protect\citeauthoryear{{Claret}}{{Claret}}{2000}]{Claret00}
{Claret}, A., 2000, A\&A, 359, 289
\bibitem[\protect\citeauthoryear{{Cody}, \& {Sasslov}}{{Cody}, \& {Sasslov}}{2002}]{Cody02}
{Cody}, A., \& {Sasselov}, D., 2002, ApJ, 569, 451
\bibitem[\protect\citeauthoryear{{Drake} et~al.}{{Drake} et~al.}{2003}]{Drake03}
{Drake}, A.J., et al., 2003, in preparation
\bibitem[\protect\citeauthoryear{{Dreizler} et~al.}{{Dreizler} et~al.}{2002}]{Dreizler02}
{Dreizler}, S., et~al., 2002, A\&A, 391, 17
\bibitem[\protect\citeauthoryear{{Guillot}, \& {Showman}}{{Guillot}, \& {Showman}}{2002}]{Guillot02}
{Guillot}, T., \& {Showman}, A.P., 2002, A\&A, 385, 156
\bibitem[\protect\citeauthoryear{{Henry} et~al.}{{Henry} et~al.}{2000}]{Henry00}
{Henry}, G.W., {Marcy}, G., {Butler}, R.P., \& {Vogt}, S.S., 2000, ApJ, 529, L41 
\bibitem[\protect\citeauthoryear{{Jenkins}, {Caldwell}, \& {Borucki}}{{Jenkins}, et~al.}{2001}]{Jenkins}
{Jenkins}, J., {Caldwell}, D.A., \& {Borucki}, J., 2002, ApJ, 564, 495
\bibitem[\protect\citeauthoryear{{Kitamura}, \& {Nakamura}}{{Kitamura}, \& {Nakamura}}{1988}]{Kitamura88}
{Kitamura}, M., \& {Nakamura}, Y., 1988, Ap\&SS, 145, 117 
\bibitem[\protect\citeauthoryear{{Mall\'en-Ornelas} et~al.}{{Mall\'en-Ornelas} et~al.}{2002}]{Ornelas02}
{Mall\'en-Ornelas}, G., et~al., 2002, astro-ph/0203218
\bibitem[\protect\citeauthoryear{{Marcy}, \& {Bulter}}{{Marcy}, \& {Bulter}}{2000}]{Marcy00}
{Marcy}, G., \& {Butler}, R., 2000, PASP, 112, 137
\bibitem[\protect\citeauthoryear{{Marsh}}{{Marsh}}{2001}]{Marsh01}
{Marsh}, T.R., 2001, MNRAS, 324, 547
\bibitem[\protect\citeauthoryear{{Mazeh}, et al.}{{Mazeh}, et al.}{2000}]{Mazeh00}
{Mazeh}, T., et~al., 2000, ApJ, 532, 55
\bibitem[\protect\citeauthoryear{{Rafert}, \& {Twigg}}{{Rafert}, \& {Twigg}}{1980}]{Rafert80}
{Rafert}, J.B., \& {Twigg}, L.W., 1980, MNRAS, 193, 79
\bibitem[\protect\citeauthoryear{{Seager}, \& {Sasselov}}{{Seager}, \& {Sasselov}}{2000}]{Seager00}
{Seager}, S., \& {Sasselov}, D.D., ApJ, 2000, 537, 926
\bibitem[\protect\citeauthoryear{{Seager}, \& {Mall\'en-Ornelas}}{{Seager}, \& {Mall\'en-Ornelas}}{2002}]{Seager02}
{Seager}, S., \& {Mall\'en-Ornelas}, G., 2002, astro\-ph/0206228
\bibitem[\protect\citeauthoryear{{Sahu}, \& {Gilliland}}{{Sahu}, \& {Gilliland}}{2002}]{Sahu02}
{Sahu}, K., \& {Gilliland}, R.L., 2002, astro\-ph/0210554
\bibitem[\protect\citeauthoryear{{Udalski} et~al.}{{Udalski} et~al.}{2002a}]{Udalski02a}
{Udalski}, A., et~al., 2002a, Acta Astron., 52, 1
\bibitem[\protect\citeauthoryear{{Udalski} et~al.}{{Udalski} et~al.}{2002b}]{Udalski02b}
{Udalski}, A., et~al., 2002b, Acta Astron., 52, 115
\bibitem[\protect\citeauthoryear{{Udry} et~al.}{{Udry} et~al.}{2002}]{Udry02}
{Udry}, S., et~al., 2002, A\&A, 390, 267
\bibitem[\protect\citeauthoryear{{von Zeipel}}{{von Zeipel}}{1924}]{Zeipel24}
{von Zeipel}, H., 1924, MNRAS, 84, 665 

\end{thebibliography}
